\documentclass[twocolumn,preprintnumbers,amsmath,amssymb]{revtex4}

\usepackage{dcolumn}
\usepackage{amstext}
\usepackage{amsmath}
\usepackage{amssymb}
\usepackage[pdftex]{graphicx}
\usepackage{subfigure}
\usepackage[T1,OT1]{fontenc} 

\begin{document}
\title{Particle self-assembly on soft elastic shells}
\author{An{\fontencoding{T1}\selectfont\dj}ela \v{S}ari\'c and Angelo Cacciuto}
\affiliation{Department of Chemistry, Columbia University\\ 3000 Broadway, MC 3123\\New York, NY 10027 }
\renewcommand{\today} 

\begin{abstract}
 We use numerical simulations to show how noninteracting hard particles binding to a deformable elastic 
shell may self-assemble into a variety of linear patterns.
 This is a result of the nontrivial elastic response to deformations of shells. The
 morphology of the patterns can be controlled by the mechanical 
properties of the surface, and can be fine-tuned by varying the binding energy of the particles. 
We also repeat our calculations for a fully flexible chain and find that the chain conformations 
follow patterns similar to those formed by the nanoparticles under analogous conditions.
We propose a simple way of understanding and sorting the different structures  and relate it to the 
underlying shape transition of the shell. Finally, we discuss the implications of our results.
  
\end{abstract}
\maketitle

\section{Introduction}
Spontaneous assembly of components into large ordered aggregates is a ubiquitous phenomenon in nature, and is observed across all length scales. Aggregation of proteins into functional nanomachines~\cite{alberts}, formation of viral capsids~\cite{watson,klug}, packing of phospholipids into biological membranes~\cite{sackmann} and assembly of colloidal particles into macroscopic photonic crystals~\cite{pine} are just a few examples of such fundamental processes. The human body itself develops as a hierarchical self-assembly of units of different sizes. Understanding the physical mechanisms driving self-assembly of biological or artificial components will shed light on critical biological processes and simultaneously holds promise for the development of materials with novel functional, mechanical, and optical properties.

The use of fluid interfaces has revolutionized the field of self-assembly~\cite{whitesides,whitesides2} by providing a universal mechanism through which arbitrary building blocks can be driven close to each other. As a result the process of self-assembly, once specific to molecular systems, has been extended to components of much
 larger size~\cite{whitesides}. The physics behind this mechanism is well known~\cite{isra,gennes}, and its manifestations are a matter of common experience, as seen for instance in the clustering of cereal on milk surface. Local deformations in the interface profile, caused by the presence of floating objects, induce long-range capillary forces. These are the result of the minimization of the total surface area of the interface that is regulated by its surface tension. By manipulating the surface tension of the interface, and by tailoring the chemistry of the building blocks, millimeter-size objects, microchips and microcomponents have been successfully self-assembled~\cite{whitesides3, hanein,sato}.
 
In this work we try to understand how elastic surfaces, which have quite different mechanical properties from
their fluid counterparts, can be used to drive self-assembly of nanoparticles.
Unlike fluid interfaces, whose molecules can freely diffuse across their area, the elements of elastic surfaces are tethered together, and respond to small deformation in a spring-like fashion. The resulting macroscopic behavior, regulated by bending and stretching energies, is characterized by strong nonlinearities~\cite{landau}. The mechanical properties of macroscopic elastic sheets have recently been the subject of intense 
scrutiny~\cite{cerda1,cerda2,cerda3,skin,russell,witten}. However, there are several microscopic artificial and naturally occurring examples of such materials, including graphite-oxide sheets~\cite{spector,wen}, graphene 
sheets~\cite{ruoff,meyer}, cross polymerized biological membranes~\cite{tundo}, 
buckypaper~\cite{endo, hodak,hall}, the spectrin-actin network forming the cytoskeleton of Red Blood Cells~\cite{branton,lei}, and very recently they have been fabricated using close-packed nanoparticle arrays~\cite{lin}.

It is easy to show~\cite{landau,witten} that the ratio between the stretching and bending energies for an arbitrary deformation of amplitude $h$ on a flat elastic sheet of thickness $t$
scales as $E_s/E_b \sim (h/t)^2$. Therefore, for sufficiently thin sheets, bending is the preferred mode of
deformation. This has a profound effect on the way elastic surfaces respond to deformations as the only stretch-free deformation involves
uniaxial bending. Skin wrinkling under applied stress~\cite{cerda2,skin} and stress focusing via d-cone formation of crumpled paper~\cite{witten} are two beautiful examples of this phenomenon.
We have recently shown how the response to deformations of elastic nanotubes can alter the
elastic properties of a flexible filament binding to it~\cite{saric}, and drive self assembly of 
colloidal particles into helical and axial lines~\cite{pamies}.
 
 In this paper we show how the elastic response to deformations of a spherical shell can be
used to self-assemble colloidal particles in a variety of patterns that only  depend on the 
mechanical properties of the system and the amount of deformation of the surface. We also show how 
a fully flexible polymer bound to the shell will spontaneously arrange to conform to similar patterns 
observed for the nanoparticles.
What makes this problem very interesting is that unlike planar or cylindrical geometries, for which there is a clear
solution to the stretch-free deformation problem, any deformation of a spherical shell will
necessarily involve stretching of the surface. The shape of the deformation minimizing the
stretching energy in this case is  therefore not immediately obvious. Furthermore, 
a buckling transition from the spherical to a faceted icosahedral   
shape is known to take place at large stretching energies~\cite{nelson}, and finally
for sufficiently small shells (or large deformations) the ratio between bending and stretching energy becomes
independent of $h$, and only depends on the radius of the sphere $R$ and its thickness $t$,
$E_s/E_b \sim (R/t)^2$~\cite{landau}. 

\section{Model}
We model the elastic surface via a standard fishnet network representation\cite{mesh}.
Each node of the network is placed to conform to the symmetry of an icosadeltahedron. 
In such surfaces, like viral shells, all but 12 nodes have a regular triangulation with six neighbors;
12 five-fold disclinations are also present as required by Euler's theorem relating number of faces, edges and faces on a spherical triangulation. The number of nodes $N_k$ describing the surface is then related to the location of 
the five disclinations, and satisfies the constraint $N_k=10(n^2+nm+m^2)+2$~\cite{geom}. Here $n$ and $m$ indicate that one must move $n$ nodes  along the row of neighboring bonds on the sphere, and then after a
turn of $120^{\rm o}$, move for $m$ extra steps. 
 
 We studied two different sphere sizes; the smaller one has $N_k=6752$ nodes and
 symmetry described by  $n=15$ and $m=15$, the larger one contains $N_k=10832$ nodes  
 with $n=19$ and $m=19$.

To impose surface self-avoidance we place hard beads at each node of the mesh.
Any two surface beads  interact via a purely repulsive truncated and shifted  Lennard-Jones potential
\begin{equation} \label{LJ}
U_{LJ}=
\begin{cases}
4\epsilon\left[ \left( \dfrac{\sigma}{r}\right)^{12}-\left(\dfrac{\sigma}{r}\right)^{6} + \frac{1}{4}\right] & \text{, $r\leq 2^{1/6} \sigma$}\cr
0 &\text{, $r>2^{1/6} \sigma$}\cr
\end{cases}
\end{equation}
where $r$ is the distance between the centers of two beads, $\sigma$ is their diameter, and
$\epsilon=100 k_{\rm{B}}T$.

We enforce the surface fixed connectivity by linking every bead on the surface to its first neighbors via a harmonic spring potential
\begin{equation} \label{spring}
U_{stretching}=K_s(r-r_B)^2
\end{equation}
Here $K_s$ is the spring constant and  $r$ is the distance between two neighboring beads. $r_B=1.23\sigma$ is the equilibrium bond length, and it is sufficiently short to prevent overlap between any two triangles on the surface even for moderate values of $K_s$.

The bending rigidity of the elastic surface is modeled by a dihedral potential between  adjacent triangles on the mesh:
\begin{equation} \label{dihedral}
U_{bending}=K_b(1+\cos\phi)
\end{equation}
where $\phi$ is the  dihedral angle between opposite vertices of any two triangles sharing an edge and  $K_b$ is the bending constant.

Colloidal particles of diameter $\sigma_c=10\sigma$ are described via 
the repulsive truncated-shifted Lennard-Jones potential
introduced in Eq.~\ref{LJ} with $\sigma\rightarrow\sigma_c$. Finally the polymer is constructed 
as a ``pearl necklace'' with $N\in[{20,45}]$ monomers of diameter also  $\sigma_m=10\sigma$.
Neighboring monomers are connected by harmonic springs as in Eq.~\ref{spring} with the equilibrium bond length $r_M=1.18 \sigma_m$
and spring constant of 120$k_{\rm B}T/\sigma^2$. Polymer self-avoidance is again enforced via the repulsive truncated-shifted Lennard-jones potential
introduced in Eq.~\ref{LJ} with $\sigma\rightarrow\sigma_m$.

The generic binding of the  polymer monomers (and the colloids) to the surface is described by a Morse potential:
\begin{equation} \label{morse}
U_{Morse}=
\begin{cases}
D_0\left( e^{-2\gamma(r-r_{MB})}-2e^{-\gamma(r-r_{MB})}\right) & \text{, $r\leq 10 \sigma$}\cr
0 &\text{, $r>10 \sigma$}\cr
\end{cases}
\end{equation}
where $r$ is the center-to-center distance between a monomer and a surface-bead,  $r_{MB}$ is bead-monomer contact distance $r_{MB}=5.5\sigma$ and
$D_0$ is the binding energy. The interaction cutoff is set at 10$\sigma$ and $\gamma=1.25/\sigma$.

We used the {\sc LAMMPS} molecular dynamics package~\cite{lammps}
with a Nos\'{e}/Hoover thermostat in the $NVT$ ensemble to study the statistical behavior of the system.
The timestep size was set to $dt=0.002\tau_0$ ($\tau_0$ is the dimensionless time)
and each simulation was run for a minimum of $5 \cdot 10^{6}$ steps.
The radii of the undeformed spherical shells are $R=29.05\sigma$ for $N_k=6752$,  and $R=34.16\sigma$ for $N_k=10832$.

\section{Results}
\begin{figure*}
 \includegraphics[width=160mm]{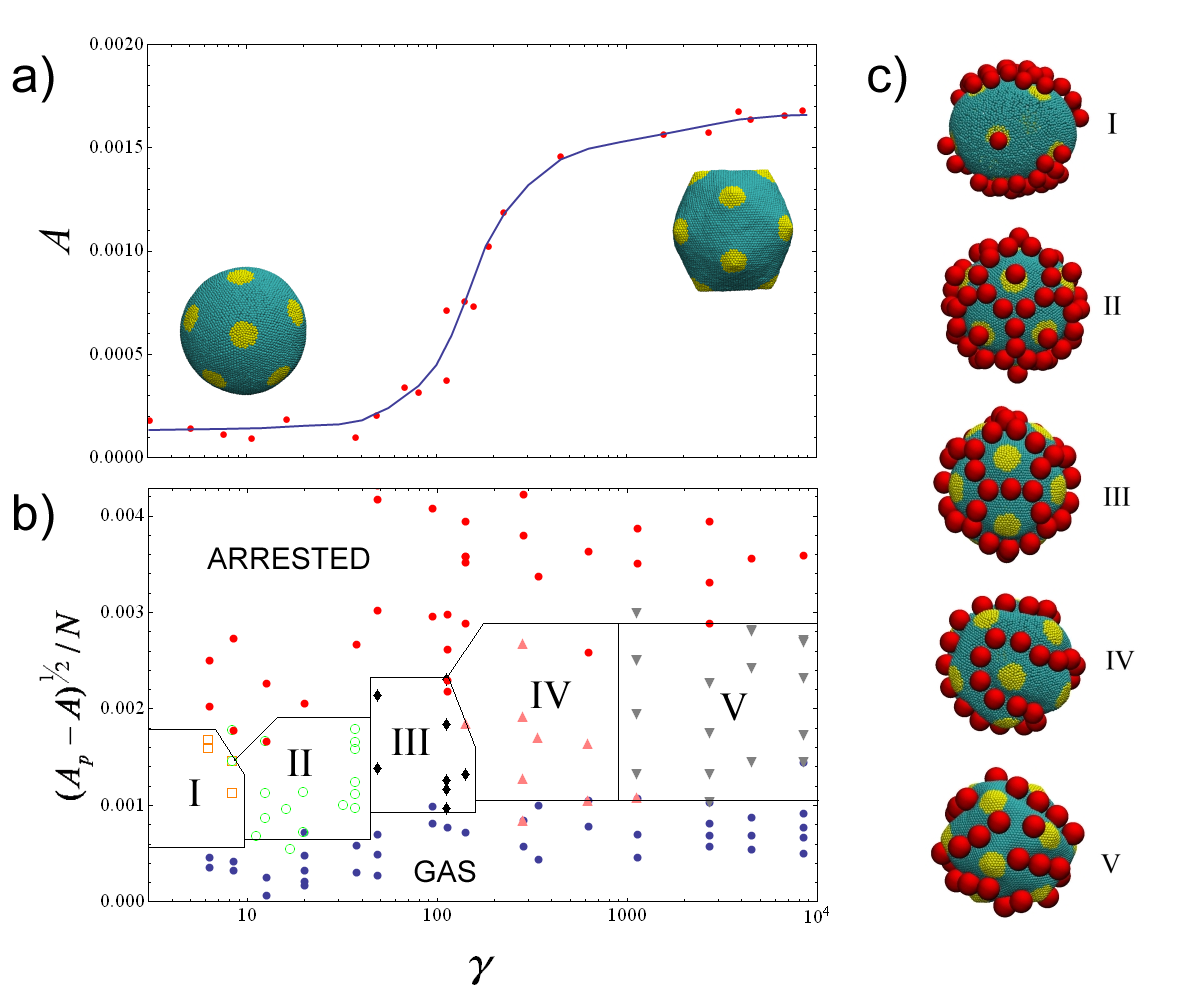}
\caption{(a) Plot of the asphericity $A$ versus Foppl-Von K\`arm\`an number $\gamma$ indicating the 
buckling transition in our model of a spherical elastic shell in the absence of colloidal particles. 
(b) Phase diagram indicating how the different aggregates formed by the colloidal particles depend 
on the mechanical properties of the shell ($\gamma$) and the degree of indentation measured
in terms of the particle-induced asphericity $(A_p-A)^{1/2}/N$ computed using $N=35$ particles. The five different phases are indicated with Roman numerals.
(c) Snapshots from our simulations of the phases indicated in the phase diagram. From top to bottom $\gamma=5.6$, $\gamma=37.5$, $\gamma=75$, $\gamma=225$ and $\gamma=4500$. For the sake of clarity, the particles defining the shell are depicted with a larger volume and the regions arund the disclinations are depicted with a lighter color.}
\label{patterns} 
\end{figure*}

To understand how  the configurational properties of the binding polymer and colloids are related to the elastic properties of the templating surface, we performed a series of simulations for many values of 
$K_s$, $D_0$ and $K_b$, and for different polymer lengths and number of colloidal particles.
 We find that a convenient way of representing our data is via the dimensionless parameter known as
 the Foppl-Von K\`arm\`an number, defined as $\gamma=YR^2/k$, where $Y$ is the Young modulus of the shell and $k$ is the bending rigidity as defined in the continuum theory of elasticity~\cite{landau}.
We begin our analysis by studying the buckling transition of the shell as a function of $\gamma$ in
the absence of binding agents. This will give us critical information about how to relate the shape of the 
templating surface (the elastic shell) and its elastic properties. To identify the buckling transition
and match the numerical parameters of our model with $\gamma$ we follow the analysis carried out in~\cite{nelson}. Fig.~\ref{patterns}A shows the results of our simulations. The surface asphericity $A$, defined as
\begin{equation} 
A=\frac{\left<\Delta R^2\right>}{\left<R\right>^2}=\sum_{i=1}^{N}\frac{(R_i-\left<R\right>)^2}{\left<R\right>^2}
\end{equation}
is plotted againts $\gamma=(4/3) K_sR^2/K_b$. The buckling transition, for which the spherical shape begins to facet, begins for values of $\gamma \gtrsim 10^2$. This result is in good agreement with that computed in ~\cite{nelson}, and represents a good test of our numerical model.
We next add colloidal particles to the system and observe their self-assembly on the surface of the 
spherical shell. Depending on the value of $\gamma$, particles arrange in patterns that minimize the
elastic energy of the shell. Figure~\ref{patterns}B shows the resulting patterns as a function of $\gamma$ 
for different values of indentation $h$, and a constant number of colloidal particles $N=35$. A convenient
way of extracting the indentation is obtained by computing  $(A_p-A)^{1/2}$, where $A_p$ is the asphericity of the 
shell in the presence of the particles and $A$ is the asphericity without them.

The general feature of the diagram is that self-assembly occurs for only a relatively narrow range of particle deformations.
When  $(A_p-A)^{1/2}$ is too small, i.e.  the surface is basically unaffected by the presence of the particles, we find no aggregation; this is the result one should indeed expect when placing $N$ noninteracting hard particles 
on the surface of a rigid sphere (al low densities). We indicate this phase as the {\it gas phase}. 
When $(A_p-A)^{1/2}$ is too large and significant deformations are induced by the 
binding particles, we find that the system becomes kinetically trapped (at least within the timescales considered in this study) into metastable states. In fact, repeating 
the simulations under the same conditions leads consistently to different and  not well defined aggregation
patterns. We indicate this phase as the {\it arrested phase}. For intermediate deformations, particles consistently self-assemble in a variety of patterns whose features are clearly related to the mechanical properties of the 
shell via $\gamma$.

For small values of $\gamma$ we  find that particles aggregate isotropically to form a two dimensional crystal on top of the sphere. The presence of these 2d crystals tends to flatten the surface underneath it. 
As a consequence, if a sufficiently large number of particles is added to the system we find that the side length of the crystal is limited by the shell diameter, and extra particles begin to aggregate on its opposite side (Fig. ~\ref{patterns}B-I).
As $\gamma$ becomes larger,  at low surface coverage, particles become localized
over the twelve disclinations, and as $N$ increases linear aggregates initially grow by linking the five fold 
disclinations on the sphere and finally form a linear network with 3-line joints  winding around the disclinations in the shell (Fig. ~\ref{patterns}B-II). 
As the templating surface begins to facet, each  segment of the network straightens revealing a clear patterns following the seams that a pentagonal tiling of the sphere would generate 
(a dodecahedron, Fig. ~\ref{patterns}B-III). It should be noticed that the previous two phases are topologically equivalent. The only difference is the 
presence of extra particles sitting on the twelve disclinations in region II of the diagram.
Increasing the surface coverage  in region III results in thickening
of the width of the dodecahedral pattern by formation of parallel and adjacent secondary lines of particles. 

At even larger values of $\gamma$, once the shell is well faceted, particles arrange into a linear
and non-connected aggregate that smoothly winds around and away from the twelve 
disclinations (Fig. ~\ref{patterns}B-IV). This pattern is
reminiscent of that of the baseball or tennis ball seam, the difference being that in our case the geometry of the aggregate is dictated by the presence of 12 topological defects, while in the baseball the seam winds around the location of the four $s=1/2$ disclinations that a thin nematic liquid crystal texture would generate on 
a sphere~\cite{vitelli,vitelli2}.

\begin{figure*}
\includegraphics[width=120mm]{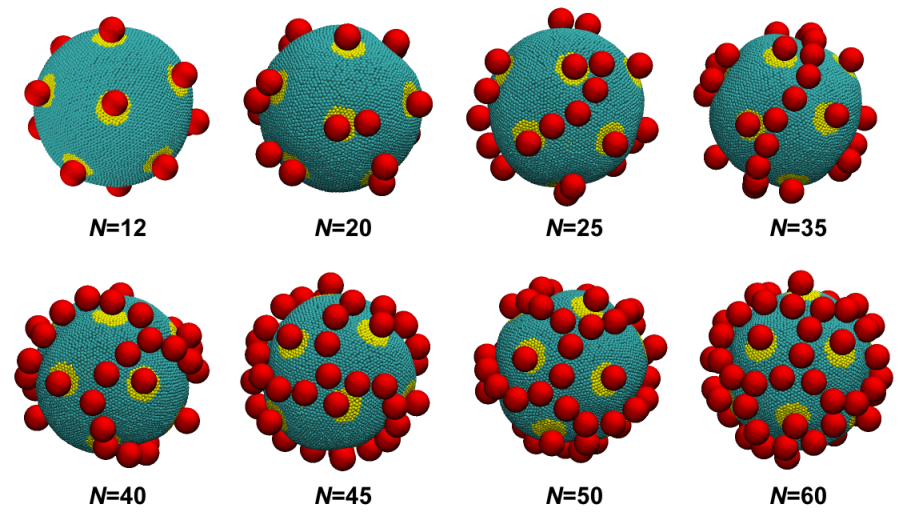}
\caption{Snapshots showing the particle aggregates as a function of surface coverage for phase II in Fig.~\ref{patterns}B, for the case $\gamma=37.5$ and $R=29.05 \sigma$.}
\label{Dynamic}  
\end{figure*}

Finally, for the largest  values of $\gamma$ we find that particles arrange into straighter but shorter
aggregates (rods) that localize into ten distinct regions of the shell. 
The length of the rods grows with the surface coverage until a critical size which depends on the 
size of the sphere and equals roughly the distance between two disclinations located at
the opposite vertices of two of the icosahedral triangles that share one edge (Fig. ~\ref{patterns}B-V).
Further increase of $N$ results in the formation of multiple rods per region 
which align parallel to each other.  Figure~\ref{Dynamic} shows the explicit dependence of the 
pattern as a function of  surface coverage in region II of the phase diagram.

\begin{figure*}
\includegraphics[width=190mm]{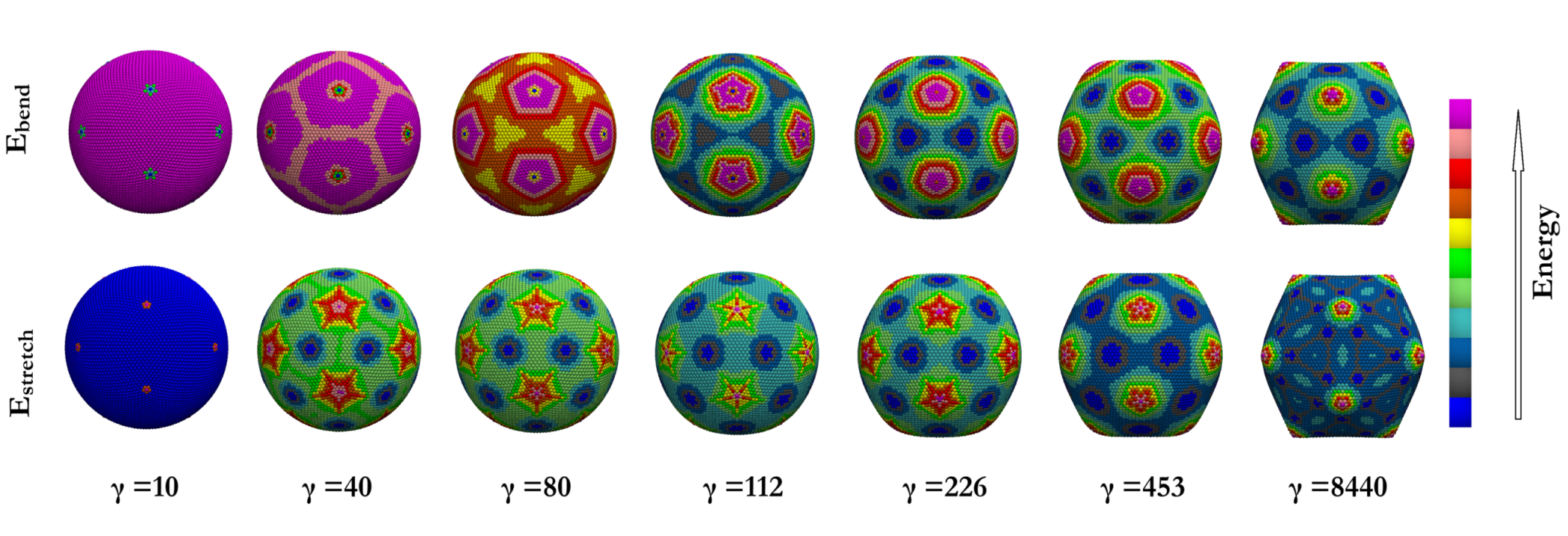}
\caption{Bending energy map (TOP) and stretching energy map (BOTTOM) 
of a spherical shell as a function of $\gamma$. Different shades indicate 
the relative strengths as indicated in the color map on the side.} 
\label{energies}  
\end{figure*}

To gain insight into the physical origin of the different patterns formed by the particles, 
we measured the strain and bending energy at each node  of the shell for different values
of $\gamma$ in the absence of the particles. Figure~\ref{energies} shows the energy map of the two contributions 
for small, intermediate and large values of $\gamma$. The results are quite revealing and provide a simple
framework from which the patterns can be understood.  Particles aggregates align to follow the low bending 
and stretching energy regions on the shell. The formation of isotropic aggregates and the presence of the particles on top of the disclinations  for small values of  $\gamma$ suggests that bending energy plays an important role in determining  the pattern in region 1 and 2 of the diagram. Regions III, IV and V are instead completely dominated by the stretching energy which is driving the shape transition of the shell.
 
Of particular interest are the two phases that occur for very large values of $\gamma$ (region IV and V). 
This condition that can be obtained either by increasing the radius of the sphere $R$ or by significantly altering the relative weight of stretching and bending energies in favor of the former. In this regime particles can easily bend the surface in regions that are far away from the icosahedral vertices, yet the only bending deformations that will not induce stretching energy are those that involve bending around a single radius of curvature. We believe that the transition from the dodecahedral arrangement  to the smooth closed loops in Fig.~\ref{patterns}B region IV is due to the large stretching energy cost that would be associated  with the formation of either sharp corners or three-lines joints.
This constraint become so severe in region V of the same figure, that the linear aggregates break 
into shorter pieces of roughly equal length that straighten by flattening the edges shared by any two triangles of the underlying icosahedral geometry.

\begin{figure*}
\includegraphics[width=120mm]{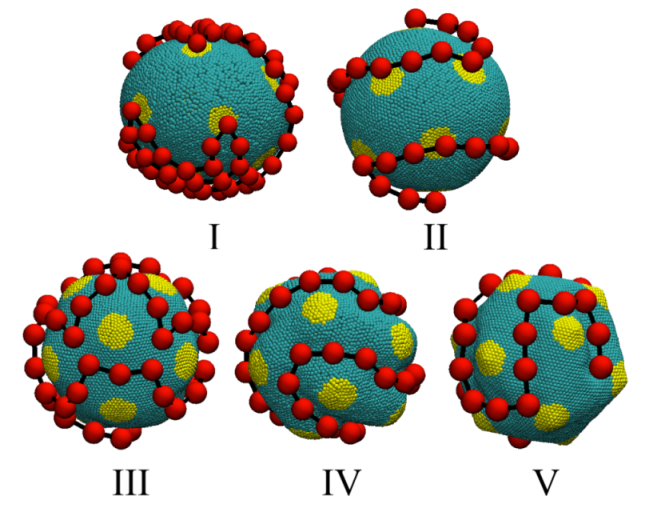}
\caption{Snapshots showing the different conformations adopted by a fully flexible chain when 
binding to a deformable elastic shell. From left to right $\gamma=5.6$, $\gamma=37.5$, $\gamma=112.5$, $\gamma=140$ and $\gamma=2700$.}
\label{patterns2}  
\end{figure*}

Interestingly, we find that almost all phases depicted in Fig.~\ref{patterns} can also be acquired by a fully 
flexible chain binding to the shell for analogous degrees of deformations and comparable number of monomers. 
The only  differences are due to the connectivity constraints on the chain.
Region II and region V  are obviously impossible to achieve with a chain; nevertheless,
the difference in region V is minimal as short rods are in this case replaced by a 
poly-line with segments having the same length of the rods formed by the colloids. 
In region II the polymer traces a simple path connecting the disclinations. 
Figure~\ref{patterns2} shows snapshots of the different phases for the fully flexible chain.
 Given these similarities between the behavior of the polymer and that of the colloidal particles, it is clear that 
the mechanism regulating the conformational changes of the chain on a flexible shell is identical to that driving the self-assembly of the colloidal particles on the same surface. 

\section{Conclusions and Discussions} 
In this paper we detail how deformable elastic surfaces can be used to mediate self-assembly of otherwise
noninteracting colloidal particles, and/or alter the conformational properties of a fully flexible chain bound to it.
We find that the structure of the aggregates (the conformation of the polymer) can be understood in terms of the 
mechanical properties of the templating shell via the Foppl-Von K\`arm\`an number $\gamma$. Specifically,
we have shown how there are two distinct regimes: one dominated by the shell bending energy,   and the other by its stretching energy. In the former case the shell acquires an overall spherical shape and particles localize on top of the disclinations and organize to link small bending energy regions. In the latter case the shell facets into an icosahedron with vertices located where the twelve five-fold disclinations reside; here particles follow low stretching pathways across  the shell.

Crucially, the underlying shell's shape transition determines the role of the disclinations in the self-assembly process. 
The twelve five folded points attract the binding colloidal particles for small values of $\gamma$ and repel them in the other regime. This result can be of great importance for controlling the 
functionalization of mesoparticles such as for instance colloidosomes~\cite{colloidosomes}. 
In fact, we have shown how a small number of particles can be localized around the twelve disclinations 
in the bending dominated regime, and around the vertices of a dodecahedron (dual to the icosahedron) 
in the stretching dominated regime.

Our results can be considered as another example or an extension of the  ideas discussed by Nelson~\cite{nelson2,vitelli} regarding the functionalizability of  the four $s=1/2$ disclinations  in a nematic liquid crystal texture covering a colloidal particle. One of the main differences, apart from the overall symmetry of the problem, is that in our case the shape of the template is allowed to change in response to the elastic strains induced by the presence of the defects, and that self-assembly (or conformational changes in a flexible chain) is driven by the elastic response to deformation of the shell. It would be interesting to study how particles self-assemble over a deformable shell under the overall
tetrahedral symmetry provided by a nematic liquid crystal texture.

\section*{ACKNOWLEDGMENTS}
This work was supported by the National Science Foundation under Career Grant No. DMR-0846426. We thank Josep C. P\`amies for helpful discussions.

\end{document}